\documentclass[12pt]{article}
\usepackage{graphics}
\begin{document}
\begin{center}
{The Self-energy of Nucleon for the Photoproduction of Pion \\ in the Low-energy Region}
\end{center}
\begin{center}
{Susumu Kinpara}
\end{center}
\begin{center}
{\it Institute for Quantum Medical Science \\ Chiba 263-8555, Japan}
\end{center}
\begin{abstract}
The scattering amplitude of the photoproduction of $\pi^{+}$ is calculated using the helicity formalism.
The pseudovector coupling $\pi$-N interaction with the non-perturbative term is applied to the perturbative expansion.
The parameters for the self-energy are taken from the results of the $\pi$-N scattering in our previous study. 
\end{abstract}
\section*{\normalsize{1 \quad Introduction}}
\hspace*{4.mm}
Nucleon is a fundamental object and plays a central role in a lot of nuclear phenomena. 
The assumption is made true by elucidating the interaction between particles and the mechanism of the reaction processes. 
To proceed the study the field theoretical method is essential and it is main purpose to examine the validity of the techniques of the calculation
by applying to phenomena known as the experimental facts.
For the nucleon is under the nuclear force and it is described by the meson-exchange model the elements of the system are mainly nucleon 
and pion which is endowed the lightest mass that is the longest range of the force. 
\\\hspace*{4.mm}
Once the interaction is prepared the calculation of the observable is made by the perturbative expansion of the T-product.
The procedure of the quantum electrodynamics is an example for the investigation of the pion-nucleon system to some extent. 
There are differences between them.
In the latter case which is the subject of the present study the interaction is the derivative type and in usual the renormalization is not applicable. 
Therefore the non-perturbative relation on the basis of the equation of the motion is valuable to draw some conclusions about the vertex part.
Since the axial-vector current is not conserved the non-perturbative term arises apart from the cancellation between the non-covariant terms
in the perturbative calculation.
It is interesting that the non-perturbative term left includes the self-energy and the divergence is removed along with the counter terms.  
\\\hspace*{4.mm}
The renormalized propagator and the vertex with the non-perturbative term makes possible to calculate any processes related to nucleon.
While the results are free from the divergences the coupling constant remains to be uncertain.
The strength is an adjustable parameter associated with the degree of the off-shell behavior 
although the on-shell condition relates the pseudovector interaction to the pseudoscalar one.
It should be allowed to shift the pseudovector coupling constant from the standard value $f\sim$ 1 
due to the effect of the non-perturbative term.
As long as the knowledge of the numerical results of us the magnitude of $f$ is required to make smaller so as to understand the experimental results.
\\\hspace*{4.mm}
In spite of the modification of the $\pi$-N interaction for preparing the renormalized propagator 
it is not useful to consider the property of the $\pi$-N coupling constant
since the self-energy does not depend on it in the present lowest-order calculation.
Then the scattering process with the external line of pion is appropriate to examine the magnitude 
and the dependence on the variables such as the energy.
The pion-nucleon scattering has two legs of pion on the diagram and
it is a basic process capable of the approach by the systematic formulation as has been shown in our previous study \cite{Kinpara}.
\\\hspace*{4.mm}
The model of the non-perturbative method is suitable for the description of the intermediate energy region.
Meanwhile the low-energy region below the $\it\Delta$-resonance needs another approach to avoid the inclusion of the higher-order terms which could
give the drastic change of the form of the self-energy.
The matrix inversion is a practical mean giving the low energy parameters of the $\pi$-N scattering in terms of the self-energy.
Our purpose of this study is to investigate the role on the photoproduction of pion.
\\
\section*{\normalsize{2 \quad The self-energy for the helicity amplitude}}
\hspace*{4.mm}
The pion-nucleon system is formulated by the lagrangian of the fields and the type of the $\pi$-N interaction is the pseudovector coupling.
It is known that the system is described by the pseudoscalar coupling too. 
The former interaction is still a controversial issue because the renormalization 
is not likely to give the finite result of the correction of the higher-order graphs for the nucleon propagator.
For the vertex part of the interaction these two types are connected by the non-perturbative term containing the part of the self-energy.
In general the form of the self-energy is given as 
\begin{eqnarray}
\Sigma(p) = M c_1(p^2) - \gamma\cdot p \, c_2(p^2)
\end{eqnarray}
in terms of the coefficients $c_i(p^2)$ ($i$=1,2) as a function of the four-momentum $p$ of nucleon.
The $M$ is the nucleon mass and the $\gamma_\mu$ is the gamma matrix.
\\\hspace*{4.mm}
Assuming the expansion of $c_i(p^2)$ in the vicinity of $p^2 = M^2$ the form of $\Sigma(p)$ is determined by preparing coefficients of the series expansion.
In our previous study the scattering parameters of the $S$-wave for the pion-nucleon elastic scattering has been used to construct $\Sigma(p)$ 
by solving the equation of the matrix inversion on four unknown coefficients up to the order of $(p^2-M^2)^2$.
The remaining two variables are dependent on the solutions and the relations between them are settled by the condition of the renormalization.
For only the $S$-wave survives in the limit of the pion momentum $\vec{p}^{\;2}\rightarrow0$ of the $\pi$-N scattering
the choice of them as the input is thought to be suitable for the description of the low-energy region.
\\\hspace*{4.mm}
The set of the output values ($c \equiv c_1^{(0)} = c_2^{(0)}\,$,$\,c_2^{(1)}\,$,$\,c_1^{(2)}\,$,$\,c_2^{(2)}\,$) is responsible for the construction of the
propagator $G(p)$ through $\alpha(p^2)$ and $\beta(p^2)$ in $G(p) = (\alpha (p^2) \gamma \cdot p + \beta (p^2) M)/(p^2-M^2)$.
It is verified that the condition of the renormalization $\alpha (M^2) = \beta (M^2) = 1$ is satisfied automatically.
The $\alpha(p^2)$ and $\beta(p^2)$ are expanded in powers of $p^2-M^2$ similar to $c_i(p^2)$.
Applying the renormalized propagator to the calculation of the amplitude 
the exact one is replaced with the approximate one $\alpha(p^2) \approx 1 + \alpha^{(1)} (p^2-M^2)$ eliminating the terms of the order $O((p^2-M^2)^2)$.
The procedure is also used to obtain $\beta(p^2)$.
\\\hspace*{4.mm}
These are expressed by the set of the coefficients and given as follows
\begin{eqnarray}
\qquad \qquad \quad \alpha^{(1)} = \frac{1}{4 M^2}\cdot \frac{c^2}{1+c}-c^{(1)}_2 +M^2(c_1^{(2)}-c_2^{(2)})   \qquad\qquad\qquad\qquad
\end{eqnarray}
\begin{eqnarray}
\beta^{(1)} = \alpha^{(1)} + \frac{1}{2 M^2}\cdot \frac{c}{1+c} \qquad\qquad\qquad\qquad\qquad
\end{eqnarray}
In the previous study the term of the $c_i^{(2)}$ has not been added \cite{Kinpara}. 
Inclusion of the term makes the numerical results worse and then the terms of the higher-order would be required to improve the results.   
Turning the scalar meson $\sigma$ off the coefficient $c$ results in $c \sim -1$ and the propagator $G(p)$ becomes too large to obtain the reasonable values 
as seen in the $\pi$-N elastic scattering.
The process mediated by the $\sigma$-meson is essential to calculate the parameters of the $\pi$-N elastic scattering. 
\\\hspace*{4.mm}
The self-energy has an effect on the vertex part to correct the scattering amplitudes for various phenomena.
In the quantum electrodynamics (QED) the vertex part is connected to the propagator and the relation is known as the Ward-Takahashi (W-T) identity.
This relation is satisfied also in the case of the nucleon interacting with the photon and the pion.
Particularly the anomalous interaction of the photon-nucleon-nucleon vertex is ascribed to the identity.
Using the value of the coupling parameter $f \sim 0.8$ and choose one of the models of the self-energy 
the simultaneous fit to the strength of the anomalous part of proton and neutron is attained.
\\\hspace*{4.mm}
Similar to the prescription of the non-perturbative relation in QED the $\pi$-N-N vertex $\Gamma(p,q)$ is found out to accompany the term as
\\
\begin{eqnarray}
\Gamma(p,q) = \Gamma_0(p,q) + G(p)^{-1} \,\gamma_5 + \gamma_5 \,G(q)^{-1}
\end{eqnarray}
\begin{eqnarray}
\Gamma_0(p,q) = \gamma_5 \,\gamma\cdot (p-q) + O((f/m)^2)
\end{eqnarray}
\\
where $\Gamma_0(p,q)$ is the proper vertex defined by the perturbative expansion 
in terms of the ratio of $f$ and the mass of pion $m$.
Hence setting the on-shell conditions $\gamma\cdot p -M \rightarrow 0$ and $\gamma\cdot q -M \rightarrow 0$ for the incoming ($q$) and the outgoing ($p$) momenta 
the lowest-order of $\Gamma(p,q)$ reduces to the vertex of the pseudoscalar interaction as $\Gamma(p,q) \rightarrow -2 M \gamma_5$
with the coupling constant $G \equiv 2 M f/m$ besides the isospin. 
\\\hspace*{4.mm}
Our interest is the calculation of the photoproduction of single pion and to examine whether the self-energy plays a decisive role or not.
The process $\gamma + p \rightarrow \pi^{+} + n$ is taken as an example.
The other types of the photoproduction of pion reaction are treated like the present case.
They are classified by the state of the isospin.
For the $\pi^{+}$ production the initial state is same as the compton scattering and the optical theorem is available.
The procedure is tractable when the cross section in the forward direction is understood quantitatively by taking into account the scattering
of photon by the cloud of the virtual pions or the photon-photon-proton-proton vertex.
Alternatively the scattering amplitude is calculated by the lowest-order perturbative expansion.
The perturbative part $\Gamma_0(p,q)$ is approximated by $\Gamma_0(p,q) \approx \gamma_5 \,\gamma\cdot (p-q)$ neglecting the higher-order corrections.
\\\hspace*{4.mm}
The perturbative expansion of the $S$-matrix is used to calculate the process of the $\pi^{+}$ production.
The Feynman diagram is seen in Ref. \cite{Schweber}.
The central part $v_{s^\prime s}(p^\prime,p)$ consists of a set of four terms put between the Dirac spinors 
of the initial proton state with the four-momentum $p$ and the spin $s$ and the final neutron state with the four-momentum $p^\prime$ and the spin $s^\prime$
\begin{eqnarray}
v_{s^\prime s}(p^\prime,p) \equiv \bar{u}^{(s^\prime)}(p^\prime) \gamma_5 \,[\, a+b\,\gamma\cdot k \, \gamma\cdot \epsilon_\chi
+ c\, \gamma\cdot\epsilon_\chi + d\, \gamma\cdot k\,]\,u^{(s)}(p)
\end{eqnarray}
\begin{eqnarray}
a = a^\prime \, q \cdot \epsilon_\chi = \frac{1}{q \cdot k} \, q \cdot \epsilon_\chi
\end{eqnarray}
\begin{eqnarray}
b = -\frac{1}{2} \,(\,\frac{1}{p \cdot k}+\alpha+\beta\,) +\frac{\kappa_p + \kappa_n}{4 M^2}\qquad\nonumber\\\nonumber\\
-\frac{1}{2 p \cdot k} [\, \kappa_p (1+\frac{p \cdot k}{2 M^2})
- \kappa_n \frac{p\cdot k}{p^{\,\prime}\cdot k} (1-\frac{p \cdot q}{2 M^2}+\frac{m^2}{4 M^2})\,]\nonumber\\\nonumber\\
-\frac{\kappa_p + \kappa_n}{2}(\alpha+\beta)-\frac{(\kappa_p-\kappa_n) \alpha}{2 M^2}(p\cdot q -\frac{m^2}{2})-\frac{\kappa_p \alpha \,q \cdot k}{2 M^2}
\end{eqnarray}
\begin{eqnarray}
c = - \,\frac{p \cdot k}{M} \,\alpha - \frac{\kappa_p -\kappa_n}{2 M}-\frac{\kappa_p + \kappa_n}{2 M}(\alpha+\beta) p \cdot k 
+\frac{\kappa_n (\alpha+\beta) \,q \cdot k}{2 M}
\end{eqnarray}
\begin{eqnarray}
d = d^{\,\prime} \, q \cdot \epsilon_\chi = \frac{\kappa_n}{2 M}\,(\,\frac{1}{p^{\,\prime}\cdot k}-\alpha-\beta\,) \, q \cdot \epsilon_\chi 
\end{eqnarray}
\\
in which $k$, $q$ are the four-momenta of photon and pion and $\epsilon_\chi$ is the polarization vector of photon 
with the subscript specifying two axes perpendicular to the direction of the photon momentum respectively.
The superscripts of $\alpha^{(1)}$ and $\beta^{(1)}$ are omitted for the sake of the brevity.
When $\alpha = \beta = 0$ the $v_{s^\prime s}(p^\prime,p)$ reduces to that of the pseudoscalar coupling.
The strength of the anomalous interactions of proton and neutron are taken from the experimental values as $\kappa_p$ = 1.79 and $\kappa_n$ = $-$1.91 respectively.
\\\hspace*{4.mm}
To connect $a \sim d$ to the elements of the helicity formalism \cite{Jacob}\cite{Walker2}
the quantization axis of the final spin state is required to be coincident with each other.
The final helicity state is transformed to the one in the usual $z$-axis by operating the $D$-function. 
The resulting scattering amplitude is a linear combination of the helicity amplitudes on the neutron spin parallel and anti-parallel to the direction
of the momentum.
Consequently the elements $A_{+}^{J} \equiv A^J$, $A_{-}^{J} \equiv B^J$, $B_{-}^{J} \equiv C^J$ and $B_{+}^{J} \equiv D^J$ for $J = l+\frac{1}{2}$ ($l \ge 0$)
are obtained by the integration of the integrand $\xi_{i\pm}$ and $\zeta_{i\pm}$ ($i$$\,$=$\,$0,$\,$1)
on the scattering angle $z ={\rm cos}\,\theta$ as
\begin{eqnarray}
A_{\pm}^{J} = \frac{A}{2} \int_{-1}^{1} d z \, (\,P_l(z) \pm P_{l+1}(z)\,) (\,\xi_{0\pm}+\xi_{1\pm} \,z\,)  
\end{eqnarray}
\begin{eqnarray}
B_{\pm}^{J+1} = \pm \sqrt{\frac{l(l+1)}{(l+2)(l+3)}} \,B_{\pm}^{J} \qquad\qquad\qquad\qquad\nonumber
\end{eqnarray}
\begin{eqnarray}
   \qquad +\frac{A}{2} \sqrt{\frac{l+1}{l+3}} \int_{-1}^{1} d z \, (\,P_l(z) - P_{l+2}(z)\,) (\,\zeta_{0\pm}+\zeta_{1\pm} \,z\,)
\end{eqnarray}
\\
in which the $P_l(z)$ is the Legendre polynomial of the $l$-th order.
The $\xi_{i\pm}$ and $\zeta_{i\pm}$ are given as
\begin{eqnarray}
\xi_{0\pm} = W - Y \pm (X-2 Z_0) 
\end{eqnarray}
\begin{eqnarray}
\xi_{1\pm} = -X \mp (2 Z_1 +W-Y)
\end{eqnarray}
\begin{eqnarray}
\zeta_{0\pm} = \mp \, \zeta_{1\pm} = -X \pm (W+Y)
\end{eqnarray}
\begin{eqnarray}
W = \frac{\omega \, q}{\sqrt{2}}\cdot\frac{c+b(E+M+\omega)}{(E^\prime+M)(E+M)} = - Z_1
\end{eqnarray}
\begin{eqnarray}
X = \frac{q^2}{\sqrt{2}}\cdot\frac{a^{\,\prime}+d^{\,\prime} (E-M+\omega)}{E^{\,\prime}+M}
\end{eqnarray}
\begin{eqnarray}
Y = - \frac{\omega \, q}{\sqrt{2}} \,\{\frac{a^{\,\prime}}{E+M}-d^{\,\prime}(1+\frac{\omega}{E+M})\} -W
\end{eqnarray}
\begin{eqnarray}
Z_0 = \frac{1}{\sqrt{2}} \{ c -b (E-M+\omega) \}
\end{eqnarray}
and the overall constant is
\begin{eqnarray}
A = \sqrt{\frac{\alpha_e \,G^{\,2} (E^{\,\prime}+M) \,q}{8\pi\,(E+M) \,\omega}} 
\end{eqnarray}
with the neutron energy $E^{\,\prime}$, the proton energy $E$ and the photon energy $\omega$ in the center of mass system ($q \equiv \vert\,\vec{q}\,\vert$).
The $\alpha_e$ is the fine structure constant.
\\
\section*{\normalsize{3 \quad The results of the calculation in the threshold region}}
\hspace*{4.mm}
When the scattering angle appears in the denominator the integral is performed for each term of the expansion series.
Each of the coefficients $a^\prime$, $b$, $c$ and $d^{\,\prime}$ is
a function of ${\rm cos} \,\theta$ only through $x \equiv q_0 -q\,{\rm cos} \,\theta$.
The property is helpful to examine the threshold regions ($q \sim 0$) where the replacement $x \approx q_0$ is appropriate
and the coefficients are independent of $z = {\rm cos}\,\theta$.
\\\hspace*{4.mm}
Under the approximation the series of the elements stops at $J\,$=$\,$3/2.
The elements remaining are given as
$A_{\pm}^{3/2} = A\, \xi_{1 \pm}/3$, $A_{\pm}^{1/2} = \pm A_{\pm}^{3/2}+A\, \xi_{0 \pm}$ and $B_{\pm}^{3/2} = A\, \zeta_{0 \pm}/\sqrt{3}$.
The other elements ($J\ge\,$5/2) become $0$.
There exists the relation $B_{\pm}^{5/2} = 0$ unexpectedly due to the cancellation between the first and the second terms of the recursion relation in Eq. (12). 
The higher components of the partial wave dropped approximately return to their original values as the energy increases.
\\\hspace*{4.mm}
For the helicity elements in the low energy region the restriction $J \le 3/2$ is allowed
and the cross section $\sigma$ of the process $\gamma + p \rightarrow \pi^{+} + n$ 
\begin{eqnarray}
\sigma = \pi \sum_{l=0}^{\infty} \,(l+1)\, (\, \vert A_{+}^J \vert^2 +\vert A_{-}^J \vert^2  +\vert B_{+}^J \vert^2 +\vert B_{-}^J \vert^2 \,)
\end{eqnarray}
consists of a several of the terms where the summation of the $B_{\pm}^{J}$ part starts from $l=1$.
At the region the elements of the $J=3/2$ part has the approximate relation known as $A^{3/2} + B^{3/2} \approx C^{3/2} + D^{3/2} \approx 0$ 
making possible to reduce the number of the terms furthermore.
It arises from the form of the sum $\sim X\sim O(q^2)$ smaller than the main terms $W$, $Y$ and $Z_i$ ($i\,$=$\,0,1$) in each element.
\\\hspace*{4.mm}
Using the $x \approx q_0$ approximation the calculation of $\sigma$ is probably applicable only to the low energy region.
The lowest-order pseudoscalar model without the anomalous interaction shows the dependence of the energy 
having the broad peak around the region below the resonance energy favorable to proceed the investigation of the correction.
Regarding to the volume it is about half as large as the experimental fact and the additional effects are necessary 
to describe the mechanism of the pion production quantitatively.
The anomalous interaction is not effective against the lack of the volume 
and makes the energy dependence of $\sigma$ change to increase gradually over the resonance region.
\\\hspace*{4.mm}
To supply the strong positive effect 
the self-energy is taken into account for the diagram containing the Dirac part of the vertex of the electromagnetic interaction.
In other words the parameters of the self-energy $\alpha$, $\beta$ and the anomalous interactions $\kappa_i \,(i=p,n)$ are considered as roughly the same order 
and are retained up to the first-order of $b$, $c$ and $d^{\,\prime}$.
The $a^\prime$ does not contain these parameters from the outset.
In the actual calculation of the cross section the $\alpha$ and $\beta$ of the second order ($\sim \kappa_i\alpha$ and $\sim \kappa_i\beta$) 
are replaced with the previous values.
They are associated with the derivation of the anomalous magnetic moment to reproduce the experimental values.
\\\hspace*{4.mm}
The cross section $\sigma$ is calculated as a function of the laboratory energy.
While the curve has the property of the increasing tendency it does not change to the decrease above the resonance region.
It means that the set of $\alpha$ and $\beta$ suitable for the region is different from 
that of the threshold region determined by the $\pi$-N scattering.
The intermediate energy region is actually expressed by the other set on the basis of the perturbative calculation with the non-perturbative term. 
In fact the theoretical value of the anomalous magnetic moment of nucleon is obtained by the vertex correction with the previous set of the self-energy.
The numerical values are small ($\alpha, \beta \le 1$) in comparison with those of the low-energy $\pi$-N scattering. 
To remove or replace the $\sim \alpha \kappa_i$ and $\sim \beta \kappa_i$ ($i=$$\,p$,$\,n$) terms takes account of the anomalous interaction
arising from the joint use of the vertex correction and the self-energy of nucleon. 
It is reasonable that the theoretical curve overestimates because of the coupling constant $f$ 
which is reduced about 20$\%$ from the standard value to adjust the magnetic moment of proton and neutron to the experimental values simultaneously 
as shown in our previous study. 
\\\hspace*{4.mm}
As the incident energy increases it is difficult to disregard the second term of $x$.
Particularly when $q \gg m$, $x \rightarrow 0$ in the forward direction.
The correction of $a^{\,\prime}$ is important compared with $d^{\,\prime}$ 
in which the denominator is $p^{\prime}\cdot k =\omega(E+\omega-x)$ and the effect is weaken by the other terms
below the intermediate energy region.
The $a^{\,\prime}$ is expanded in powers of $z\,q/q_0$ as $a^{\,\prime}\,=\,\omega^{-1}\,q_0^{-1}+q \,\omega^{-1}\,q_0^{-2}\,z+O(z^2)$.
The correction $\delta\,a^{\,\prime} = q \,\omega^{-1}\,q_0^{-2}\,z$ is important since it is comparable to $\omega^{-1}\,q_0^{-1}$ at $q_0 \sim q$.
\\\hspace*{4.mm}
The present calculation achieves to reach the volume of $\sigma$ at the laboratory energy $\sim 0.3 \,{\rm GeV}$ and enables us to study the connection to 
the $\Delta (1232)$ resonance.
The application to the helicity states is interesting to examine the branching ratio observed by the experiment.
In favor of the time reversal invariance the amplitude $T_\lambda$ of the decay from the $\Delta (1232)$ resonance to the helicity $\lambda$ state is given as
$T_{1/2} \propto A_{+}^{3/2}-A_{-}^{3/2}$ and $T_{3/2} \propto B_{-}^{3/2}-B_{+}^{3/2}$ excluding the common factor.
The overall phase is arbitrarily determined by employing the extra phase in the polarization vector.
\\\hspace*{4.mm}
Due to the relation $Z_1 = -W$ in Eq. (16) the ratio is determined to be $T_{3/2}\,/\,T_{1/2} = -\sqrt{3}$
irrespective of the detail of the amplitude under the $x \approx q_0$.
For the $\lambda = 3/2$ the branching ratio is $\vert T_{3/2} \vert^2/(\vert T_{1/2} \vert^2+\vert T_{3/2} \vert^2) = 0.75$ 
a little smaller than the experimental value 0.78 -- 0.79 \cite{PDG}.
The $\delta\,a^{\,\prime}$ correction makes the value change from 0.75 to 0.84 by breaking the simple relation between the elements.
Recently it has been verified that the use of the $\sim \alpha \kappa_i$ and $\sim \beta \kappa_i$ terms with the $\alpha$ and $\beta$
obtained from the method of the matrix inversion for the $\pi$-N scattering 
yields the numerical value 0.78 up to the $O(z^3)$ order of the $\delta\,a^{\,\prime}$ correction.
\\
\section*{\normalsize{4 \quad Summary and remarks }}
\hspace*{4.mm}
The strength of the anomalous interaction is not only a parameter and it has a dynamical origin related to the vertex correction by the pion propagator.
The pseudovector coupling constant is adjusted because of the non-perturbative term which creates the self-energy 
appropriate to the anomalous terms.
In order to construct the photoproduction of pion from the threshold to the resonance region the amplitude needs the self-energy of nucleon 
determined by the method of the matrix inversion with the $\sigma$ meson exchange process.
The dispersion theoretical method would be effective along with the previous models
of the self-energy for the $\pi$-N scattering to understand the decrease of the cross section above the resonance region
in addition to the perturbative calculation for the intermediate region.
\\
\hspace{4.mm}

\end{document}